\documentclass[aps,prl,reprint,groupedaddress]{revtex4-1}

\usepackage{graphicx}
\usepackage{dcolumn}
\usepackage{bm}
\usepackage{amsmath}
\usepackage{makeidx}
\usepackage{amsfonts}
\usepackage{subfigure}
\usepackage{epsfig}
\usepackage{pst-grad} 
\usepackage{pst-plot} 
\usepackage{xcolor}

\begin{document}

\title{Mesoscale Structure of Chiral Nematic Shells}

\author{Ye Zhou}
\affiliation{Institute for Molecular Engineering, The University of Chicago, Chicago, Illinois 60637, United States}
\author{Ashley Guo}
\affiliation{Institute for Molecular Engineering, The University of Chicago, Chicago, Illinois 60637, United States}
\author{Rui Zhang}
\affiliation{Institute for Molecular Engineering, The University of Chicago, Chicago, Illinois 60637, United States}
\author{Julio C. Armas-Perez}
\affiliation{Institute for Molecular Engineering, The University of Chicago, Chicago, Illinois 60637, United States}
\author{Jos\'e A. Mart\'inez-Gonz\'alez}
\affiliation{Institute for Molecular Engineering, The University of Chicago, Chicago, Illinois 60637, United States}
\author{Mohammad Rahimi}
\affiliation{Institute for Molecular Engineering, The University of Chicago, Chicago, Illinois 60637, United States}
\author{Monirosadat Sadati}
\affiliation{Institute for Molecular Engineering, The University of Chicago, Chicago, Illinois 60637, United States}
\author{Juan J. de Pablo}\email{depablo@uchicago.edu}
\affiliation{Institute for Molecular Engineering, The University of Chicago, Chicago, Illinois 60637, United States and Argonne National Laboratory, Argonne, Illinois 60439, United States. }

\begin{abstract}
There is considerable interest in understanding and controlling topological defects in nematic liquid crystals (LCs). Confinement, in the form of droplets, has been particularly effective in that regard. Here, we employ the Landau--de Gennes method to explore the geometrical frustration of nematic order in shell geometries, and focus on chiral materials. By varying the chirality and thickness in uniform shells, we construct a phase diagram that includes tetravalent structures, bipolar structures (BS), bent structures and radial spherical structures (RSS). It is found that, in uniform shells, the BS-to-RSS structural transition, in response to both chirality and shell geometry, is accompanied by an abrupt change of defect positions, implying a potential use for chiral nematic shells as sensors. Moreover, we investigate thickness heterogeneity in shells and demonstrate that non-chiral and chiral nematic shells exhibit distinct equilibrium positions of their inner core that are governed by shell chirality $c$.
\end{abstract}
\maketitle
\section{Introduction}
Bulk cholesteric liquid crystals (chiral nematic liquid crystals) exhibit a variety of intriguing twisted mesophases \cite{kitzerow2001chirality}. Such twisted structures can have a single uniform helical axis and reflect light selectively, thereby making cholesteric liquid crystals prospective candidates for optical devices including displays, lasers, waveguides, and resonators \cite{muvsevivc2014integrated, humar20103d}. When surface confinement is imposed, the twisted ground state undergoes additional deformations due to geometrical frustration of chiral order \cite{bezic1992structures, xu1997chiral, sevc2012geometrical, geng2013liquid}. Such deformations are governed by a delicate balance between elasticity, chirality, and surface interactions; a myriad striking topological defect patterns are seemingly possible, such as free-standing knots \cite{sevc2014topological, orlova2015creation}, skyrmion lattices in channels \cite{fukuda2011quasi}, or strained blue phases in droplets \cite{wright1989crystalline, bukusoglu2015stimuli, martinez2015blue}. These defect patterns are of interest in the study of symmetry and topological entities and, once understood, could find potential applications in multiple disciplines.
When liquid crystals are confined within two spherical surfaces, they form curved channels that are referred to as nematic shells. In 1992, Lubensky and Prost predicted a spherical vesicle structure with four disclination lines at the vertices of a tetrahedron  \cite{lubensky1992orientational}, in contrast to configurations found in droplets or channels. Later, Nelson proposed the creation of nematic shells coated with colloids functionalized as to provide a 4-fold valence, resembling $sp^3$ hybridized chemical bonds \cite{nelson2002toward}. Using double-emulsion techniques, experiments were able to identify not only tetravalent structures, but also bivalent and trivalent LC shells with interfacial parallel orientations (planar anchoring) \cite{fernandez2007novel, lopez2011frustrated, koning2015spherical}. Through addition of surfactants, the anchoring type of either the inner or the outer surface of a shell can be tuned, leading to possible control over distinct structures \cite{lopez2011drops, lopez2012defect, lopez2009topological}. Numerical studies have also sought to investigate the role of shell geometry, elastic anisotropy and thermodynamic conditions on defect configurations using Ginzburg--Laudau minimizations and Monte Carlo methods \cite{vitelli2006nematic, shin2008topological, sevc2012defect, seyednejad2013confined, wand2015monte}. Recent work reported reported results for shells with embedded micro-particles and examined the configurations resulting from interactions between topological defects and micro-particles \cite{gharbi2013microparticles}.

In this work we consider two questions that, to our knowledge, have only been partially addressed. The first is, what configurations will the helical structure adopt when confined in shells? Second, we ask whether that chiral order can be manipulated and, if so, what are the resulting defect structures? To do so, we rely on a Landau--de Gennes (LdG) continuum model that enables systematic study of micrometer-sized nematic shells with strong parallel anchoring (degenerate planar anchoring) on both surfaces. We begin by validating our methods by comparing predictions for non-chiral nematic LC shells to available experiments. We then introduce chirality in uniform shells and examine the effects of thickness. Our results reveal the existence of a tetravalent structure, a bipolar structure, a bent structure and radial spherical structure (RSS). To characterize the phase boundaries between these structures, we introduce the shell chirality $c$ as an order parameter; it is given by the ratio of shell thickness to pitch. Importantly, in uniform shells, the transition from bipolar structure (BS) to RSS occurs in a narrow window of both chirality and shell geometry, implying that shells could provide a sharp response to external stimuli. We also investigate the role of elastic forces on the inner drop for intermediate and high values of chirality, and find that asymmetric stable configurations are highly dependent on shell chirality $c$.

\section{Model and Methods}
Our calculations rely on a Landau--de Gennes (LdG) continuum model for the tensor order parameter $Q$, defined by $Q_{ij} = S(n_in_j -\frac{1}{3}\delta_{ij}$) where $n_i$ are the $x$, $y$, $z$ components of the local director vector and $S$ is the scalar order parameter. \cite{de1993physics} The scalar order parameter $S$ is given by an ensemble average of the second Legendre polynomials evaluated for the dot product between the molecular orientations and the director. The value of $S$ varies from $-\frac{1}{2}$ to $1$. A value of $S = -\frac{1}{2}$ corresponds to a layer of molecules parallel to a flat surface, $S = 0$ corresponds to a completely disordered isotropic phase, and $S = 1$ corresponds to a perfectly ordered material. \cite{kleman2007soft} The total free energy of the system is given by
\begin{equation}
\begin{split}
f=&\int_{\mathrm{bulk}}\left(\frac{A}{2}\left(1-\frac{U}{3}\right)Q_{ij}Q_{ji} -\frac{AU}{3}Q_{ij}Q_{jk}Q_{ki} \right. \\
&\qquad\left.+\frac{AU}{4}(Q_{ij}Q_{ji})^2 \right) \,dV \\
+&\int_{\mathrm{bulk}}\left(\frac{L}{2}\frac{\partial Q_{ij}}{\partial x_k} \frac{\partial Q_{ij}}{\partial x_k} + 2q_0L\epsilon_{ikl} Q_{ij}\frac{\partial Q_{lj}}{\partial x_k}\right)\,dV \\
+&\int_{\mathrm{surf}}(W\left(\tilde Q_{ij} - \tilde Q_{ij}^\bot \right)^2)\,dS.
\end{split}
\label{free energy}
\end{equation}
\noindent where A and U are material constants. Parameter \emph{L} represents the elastic constant. We note that the one-constant approximation adopted here (the three basic deformation modes, splay, twist and bend are penalized equally) only applied to materials whose Frank elastic constants are not to dissimilar from each other. The inverse pitch is denoted by $q_0 = 2\pi/p_0$; it quantifies the system's chirality (for a non-chiral system, $q_0 = 0$), and $\epsilon_{ikl}$ is the Levi-Civita tensor. The anchoring strength is denoted by $W$; it ranges from $10^{-7}$ to $10^{-3}$ $\mathrm{J/m^{2}}$ for the systems considered in our work, which are typical of thermotropic liquid crystal-water interfaces. The elements of the $Q$ tensor are given by $\tilde Q_{ij}=Q_{ij} + \frac{1}{3}S_{eq}\delta_{ij}$; $\tilde Q_{ij}^\bot$ is the projection of $\tilde Q_{ij}$ onto the surface, defined by a surface normal $\nu_{i}$ as $\tilde Q_{ij}^\bot = P_{ik}\tilde Q_{kl}P_{lj}$. The projection operator is given by $P_{ij} = \delta_{ij} - \nu_{i}\nu_{j}$.
The first term in Eq. \ref{free energy} represents the short range contributions to the free energy (or phase free energy). That term controls the equilibrium value of the nematic order parameter through $S_{eq} = \frac{1}{4}\left(1+3\sqrt{1-\frac{8}{3U}}\right)$. The second term represents the elastic free energy, which governs long-range director distortions and penalizes elastic deformations from the helical phase in the undistorted bulk. The last term represents the surface energy, which quantifies deviations from planar degenerate anchoring on both the inner and outer boundaries of the shells. An iterative Ginzburg-Landau relaxation technique with finite differences on a cubic mesh (with a resolution of $7.15$ nm) is implemented to minimize the total free energy. \cite{ravnik2009landau}

To characterize the fine structure of the defects, we rely on a splay-bend parameter $S_{\mathrm{SB}}$ constructed from the second derivatives of the order parameter tensor $S_{\mathrm{SB}} = \frac{\partial ^2 Q_{ij}}{\partial x_i \partial x_j}$. \cite{vcopar2013visualisation} Assuming that there is no variation of the scalar order ($S = S_{eq}$), in the director field representation $S_{\mathrm{SB}}$ is given by
\begin{equation}
 S_{\mathrm{SB}} = \frac{3S_{\mathrm{eq}}}{2}\nabla(\textbf{n}(\nabla\cdot\textbf{n}) - \textbf{n}\times\nabla\times\textbf{n})
\label{eq:sb}
\end{equation}
\noindent The two terms in Equation \eqref{eq:sb} are related to the splay and bend deformations in the Frank--Oseen director representation. \cite{kleman2007soft} Large positive values of $S_{\mathrm{SB}}$ indicate a pronounced splay deformation, and large negative values indicate a significant bend deformation.

The numerical parameters employed here are used: $A = 1.17 \times 10^5 \,\mathrm{J/m}^3$, $U = 5$, $L = 6 \times 10^{-12}$ N, W = $10^{-3}$ $\mathrm{J/m^{2}}$, the radius of outer sphere is $R = 1\,\mathrm{\mu m}$.
Different initial conditions are used, including random configurations, uniform configurations and radial spherical configurations \cite{bezic1992structures}.

In our experiments, LC shells were produced using a glass capillary microfluidic device, based on the design developed by Utada and his coworkers. \cite{utada2005monodisperse} The LC 4'-pentyl-4-cyanobiphenyl (5CB, Sigma Aldrich) is co-injected with an inner aqueous phase and flow-focused by an outer aqueous phase, producing spherical shells of 5CB that encase the inner fluid and are suspended within the outer fluid. Both inner and outer aqueous phases consist of 5 wt.\% polyvinyl alcohol (PVA, MW: 13,000-23,000, 87-89\% hydrolyzed, Sigma Aldrich), which promotes planar anchoring on both shell surfaces. Shells are collected in a petri dish containing a bath of 5 wt.\% PVA before observation using a polarized-light microscope (Olympus BX51).

\section{Results and Discussion}
\subsection{Non-chiral Nematic Shells}

\begin{figure*}[h]
\centering
\includegraphics[width=12cm]{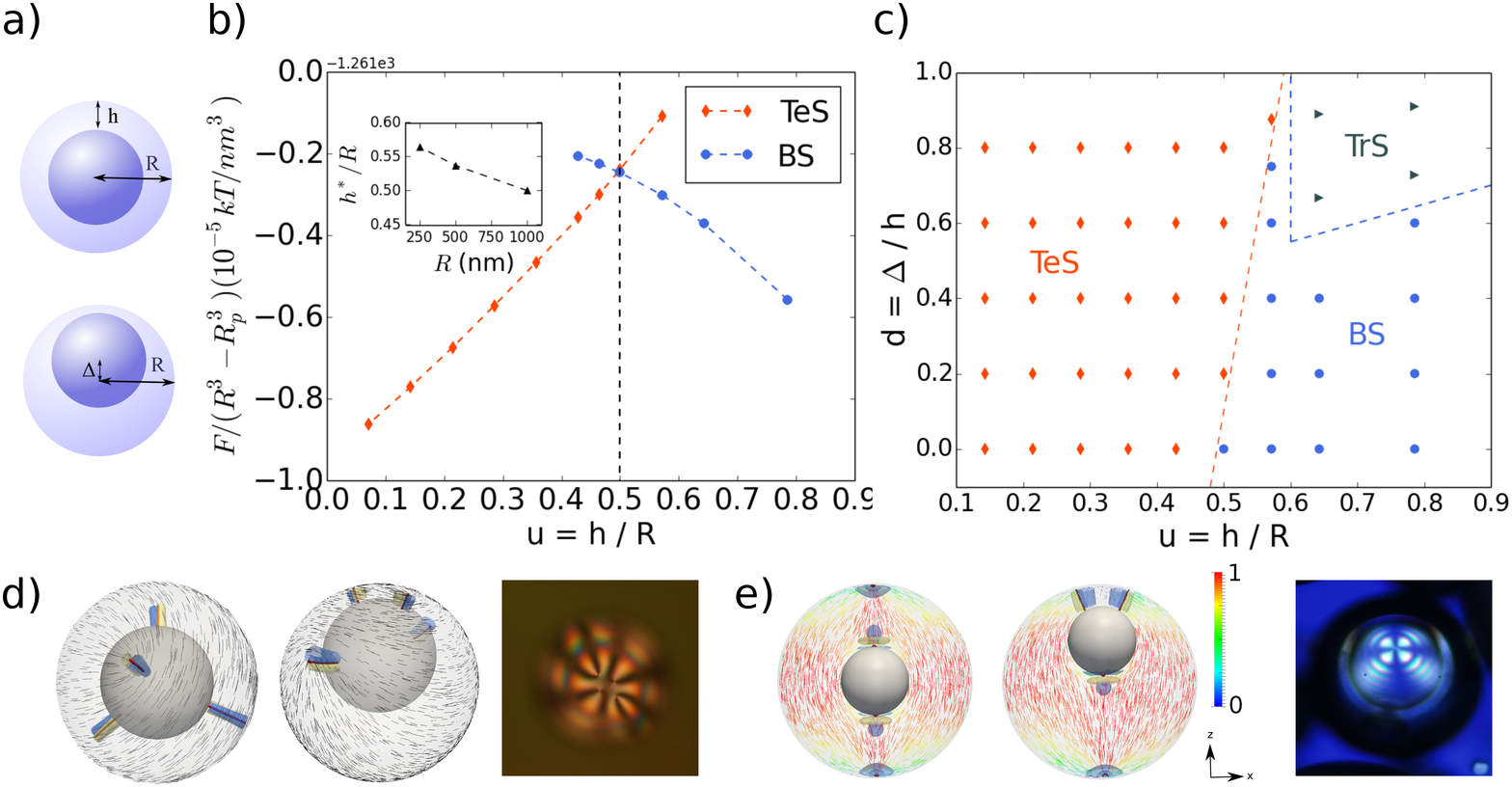}
\caption{(color online) a) Schematic diagrams for $h$, $R$ and $\Delta$. b) Free-energy density graph as a function of $u$ for bipolar structure (BS) and tetravalent structure (TeS). The cross-over of two curves demonstrates BS-to-TeS transition at $u^* \simeq 0.5$. The inset graph shows the dependence of transition thickness $u^*$ on the droplet radius $R$. c) $d$-$u$ phase diagram for non-chiral nematic shell. Red diamond, blue circle and black triangular correspond to TeS, BS and trigonal structure (TrS). d) Left and middle: simulation results of tetravalent structures (TeS) with $u = 0.429$ (left: $d = 0$, middle: $d = 0.667$). Director fields on the outer surface are shown in black lines. Right: cross-polarized light image of TeS configuration observed in experiments.  e) Left and middle: bipolar structures (BS) with $u = 0.643$ (left: $d = 0$, middle: $d = 0.667$). The director fields are on $x$-$z$ plane, colored by the projection to $z$-axis. Right: cross-polarized light image of BS configuration. For both d) and e), the defects are shown in red (isosurface for $S=0.5$); the splay and bend elastic distortions are shown in blue ($S_{\mathrm{SB}} > 0.002$) and in yellow ($S_{\mathrm{SB}} < -0.002$), respectively.}
\label{fig:N0size}
\end{figure*}

To validate our calculations, we briefly consider non-chiral nematic shells. We investigate concentric (symmetric or uniform) nematic shells with different shell thickness $h = R - R_p$, where $R_p$ is the radius of the inner drop (Fig. \ref{fig:N0size}a). For convenience, we introduce a dimensionless parameter $u = h / R$ to describe thickness, ranging from 0 to 1. In agreement with experimental observations \cite{fernandez2007novel} and past theoretical work \cite{seyednejad2013confined, vitelli2006nematic}, two different configurations are observed: a bipolar structure (BS) with two ($s = +1$) antipodal point defects, and a tetravalent structure (TeS) with four ($s = +1/2$) disclination lines. Both structures have a total topological charge of ($s = +2$) on the outer surface, as required by the Poincare-Hopf theorem \cite{poincare1885j}. Fig. \ref{fig:N0size}b shows that BS and TeS are stable configurations for thicker shells ($h \ge R/2$) and for thinner shells ($h \le R/2$), respectively. The critical thickness of this BS-to-TeS transition $u^*$ is $0.5$, same as that observed in experiments \cite{fernandez2007novel}. As illustrated in the inset of Fig. \ref{fig:N0size}b, $u^*$ depends on shell size $R$ (or curvature $1/R$). Note that a larger value of $u^* = 0.666$ was reported in previous work \cite{seyednejad2013confined}; we attribute the difference to the small (and more relevant) shell sizes considered here ($R \sim 100$ nm); indeed, our micrometer-sized simulations yield results in quantitative agreement with experiment.

Due to buoyancy effects (arising from the mismatch of density) and elastic forces induced by the liquid crystal, the inner drop may not remain in the center of the large drop. We therefore also consider briefly the role of elasticity in producing eccentric (asymmetric or inhomgeneous) shell configurations. As illustrated in Fig. \ref{fig:N0size}a, the displacement $\Delta$ is defined as the distance between the inner and outer drop centers. We also define the degree of asymmetry as $d = \Delta / h$, which varies from $0$ to $1$: $d = 0$ corresponds to a uniform shell and $d = 1$ corresponds to a shell configuration in which the inner drop touches the outer drop surface. Fig. \ref{fig:N0size}c depicts the $d$-$u$ phase diagram, consisting of three major phases: a tetravalent structure (TeS), a bipolar structure (BS), and a trigonal structure (TrS). For thin shells, as eccentricity increases (Fig. \ref{fig:N0size}d), the four disclination lines in TeS move towards the thinner layer of the shell. The trajectory and configurations for this TeS deformation in asymmetric shells have been studied in previous work \cite{sevc2012defect}. As for thicker shells, when $d$ becomes large (generally $d \ge 0.6$), BS transforms into a trigonal structure (TrS) with two ($s = +1/2$) disclination lines and a pair of point defects ($s = +1$), as shown in Fig. \ref{fig:N0size}e. In order to support our calculations, we performed experiments with 5CB to fabricate both thin and thick nematic shells with parallel anchoring (Fig. \ref{fig:N0size}d and \ref{fig:N0size}e). The agreement between simulation and experiments serves to demonstrate that the model adopted here can capture key features of nematic shells. To conclude, for the entire range of $u$ considered here, the free energies are minimized when the inner drop approaches the outer drop periphery ($d$ reaches $1)$, in agreement with our observations and also with past experimental and simulation reports \cite{fernandez2007novel, sevc2012defect, seyednejad2013confined}.

\subsection{Chiral Nematic Shells}
Previous studies of chiral nematic droplets have revealed that, when confined in a spherical geometry, nematic materials adopt new configurations, such as the twist bipolar structure and RSS. Such structures are considerably different from the helical phase observed in the bulk \cite{bezic1992structures, sevc2012geometrical, xu1997chiral}. When a chiral nematic is bounded between two spherical surfaces, in shells, we anticipate that more configurations will arise.

\begin{figure*}[h]
\centering
\includegraphics[width=12cm]{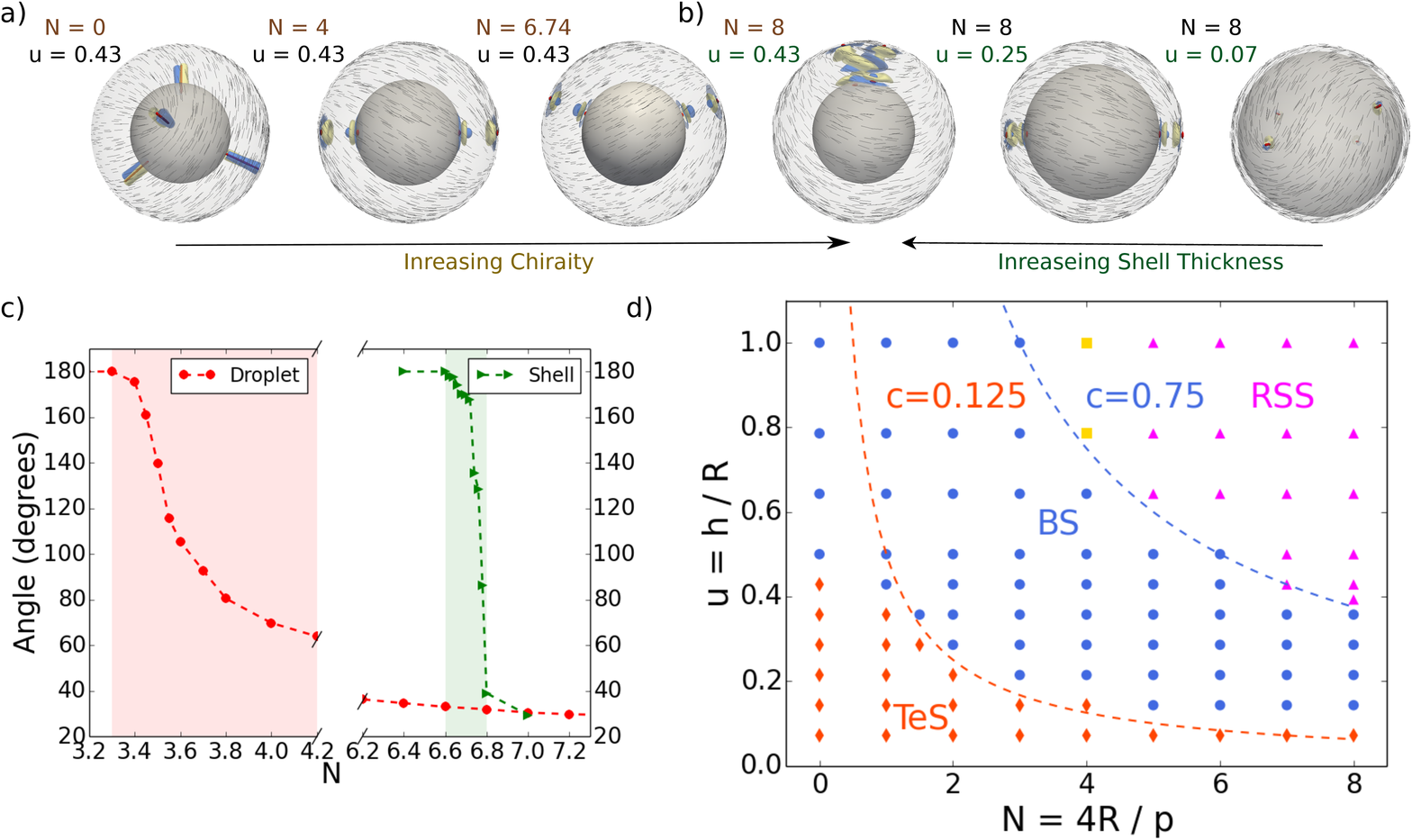}
\caption{(color online) a) and b) Shell configurations (from left to right: TeS, BS, Bent structure, RSS, BS and TeS) with different chiralities ($N =  4R/p$) and shell thicknesses ($u =  h/R$). Director fields on the outer surface are shown in black lines. Defects are in red. The splay and bend elastic distortions are shown in blue ($S_{\mathrm{SB}} > 0.002$) and in yellow ($S_{\mathrm{SB}} < -0.002$), respectively. c) Graph of $\Theta$ as a function of $N$ for droplet (red circle) and uniform shell with ($u = 0.43$) (green triangle). d) The $u$-$N$ phase diagram for concentric shells. Red diamond, blue circle, yellow square and magenta triangle correspond to TeS, BS, bent structure (BeS) and RSS, respectively. $c = h / p =  uN / 4$ represents the shell chirality.}
\label{fig:phase}
\end{figure*}

To investigate the effect of chirality on shells, we first consider high chirality concentric shells ($\Delta = 0$ and $N = 4R/p = 8$) of varying thickness $u$ (as shown in Fig. \ref{fig:phase}b). In analogy to non-chiral nematic shells, a phase transition from TeS to twisted bipolar structure (BS) occurs as the shell thickness ($u$) increases. Note that we use BS as the abbreviation for both (non-chiral) bipolar structures and (chiral) twisted bipolar structures. In highly chiral systems, we observe a second transition from BS to a radial spherical structure (RSS) shell when $u$ reaches $0.43$, and into an RSS droplet in the limit of $u = 1$. The RSS shell has concentric cholesteric layers with two surface point defects on the outer and inner surfaces, respectively. The two cholesteric $\lambda^{+1}$ disclination lines, shown in yellow in Fig. \ref{fig:phase}b ($u = 0.43$), start from the inner surface defects, and then span the shell in a helical manner until they reach the outer surface defects. \cite{sevc2012geometrical} 

As discussed in the preceding section, the non-chiral nematic shell ($N = 0$) with $u = 0.43$ exhibits a tetravalent structure (TeS). We therefore examine how, in a shell with $u = 0.43$, does TeS ($N = 0$) transform into RSS ($N = 8$) as chirality increases. Figure \ref{fig:phase}a shows two structural transitions:  a TeS-to-BS transition ($0 \le N \le 1$), followed by a BS-to-RSS transition ($6 \le N \le 7$) with the bent structure (BeS) as an intermediate configuration (Fig. \ref{fig:phase}a: $N = 6.74$). This BS-to-RSS transition, as well as the BeS structure in a chiral LC droplet, were studied in a recent publication from our group.  \cite{zhou2016structural} A comparison of BS-to-RSS transitions between these two distinct geometries is displayed in Fig. \ref{fig:phase}c. Here we define the angle between the two lines connecting outer surface point defects to the droplet or shell center as $\Theta$. Thus, $\Theta = 180^{\circ}$ for BS, $\Theta \sim 30^{\circ}$ for RSS and $\Theta$ in between for BeS. We have recently reported on a drastic decrease of $\Theta$ in chiral LC droplets within a narrow window of chirality ($3.3 \le N \le 4$), which correspond to a continuous transition from BS to RSS. Strikingly, this transition is even sharper in shells ($u = 0.43$), indicating a higher sensitivity of shell geometries to environmental disturbance (chirality) (and better performance for potential uses in sensing).

A $u$-$N$ phase diagram is shown in Fig. \ref{fig:phase}d. Three distinct phases are involved: TeS in thinner shells with low chirality, RSS in thicker shells with high chirality, and BS in between. Even though $N =  4R / p$ is commonly employed to describe chirality in droplets, it fails to characterize an extremely important parameter, thickness, in shells. This phase diagram suggests that an important parameter for shells is $c = h / p = uN / 4$, which measures the ratio of shell thickness $h$ to pitch $p$. Different from $N$, which quantifies the approximate number of $\pi$ turns along the diametrical axis, $c$ captures the number of $2\pi$ turns along the radial direction in a uniform shell.  The two dotted curves ($c =  0.125$ and $c = 0.75$) in Fig. \ref{fig:phase}d define two phase transition boundaries. In concentric shell geometries, as shell chirality $c$ increases, a TeS is initially the most stable configuration, followed by a BS, which finally transforms into RSS. Similarly to previous findings for chiral LC droplets in our group \cite{zhou2016structural}, these transitions are the result of an interplay between elasticity and chirality.

Note that the $c = 0.125$ curve deviates from the real phase boundary near $N = 0$. This is because when $N$ is close to $0$, shell chirality $c = uN / 4$ becomes too small to reveal the effect of $u$ (or $h$), thereby failing to characterize the phase boundary. Another issue for $c$ is that it only takes into account the shell thickness $h$, but fails to quantify curvature ($1/R$). To illustrate the importance of curvature in a shell geometry, we recapitulate the dependence of the TeS-to-BS transition thickness $u^*$ on curvature $1/R$ in Fig. \ref{fig:N0size}b. In this work, we simply fix $R$. A more complex expression, consisting of $R$, $Rp$ and $p$, could be proposed to provide a better order parameter in systems of varying size.

\begin{figure*}[h]
\centering
\includegraphics[width=12cm]{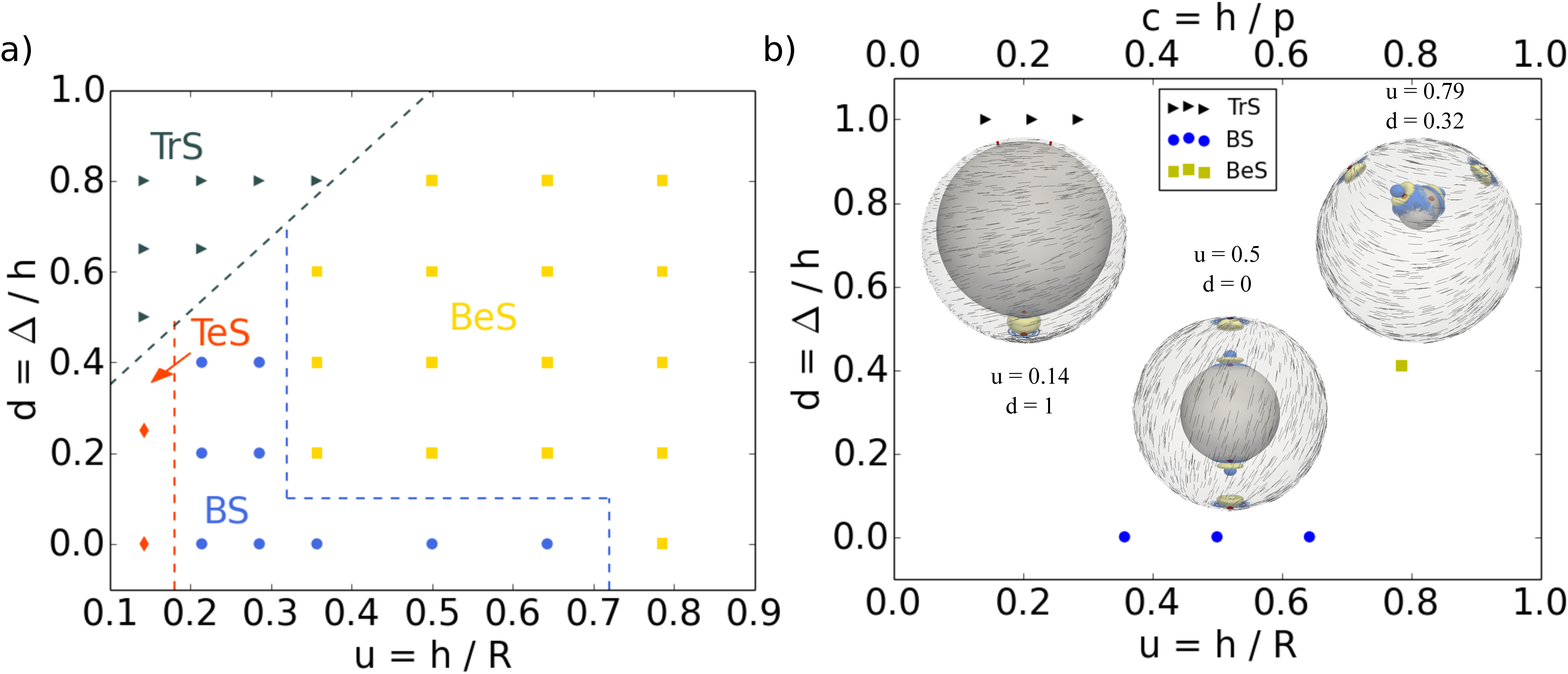}
\caption{(color online) a) $d$-$u$ phase diagram for chiral nematic shell ($N = 4$). Red diamond, blue circle, black triangle and yellow square correspond to TeS, BS, TrS and bent structure (BeS), respectively. b) Diagram of the relative energetically favorable position $d$ for different shell thickness. Inset images are representative shell configurations for TrS ($u = 0.14$, $d =1$), BS ($u = 0.5$, $d = 0$) and BeS ($u = 0.79$, $d = 0.32$). Director fields on the outer surface are shown in black lines. Defects are in red. The splay and bend elastic distortions are shown in blue ($S_{\mathrm{SB}} > 0.002$) and in yellow ($S_{\mathrm{SB}} < -0.002$), respectively.}
\label{fig:N4pos}
\end{figure*}

To the best of our knowledge, the role of elasticity on inner drop position in chiral nematic shells has been barely studied \cite{wand2015monte}. Here we consider two representative cases: intermediate chirality ($N = 4$) and high chirality ($N = 8$). Figure \ref{fig:N4pos}a shows the $d$-$u$ phase diagram for chiral nematic shells with $N = 4$. It exhibits three essential differences from what is observed for $N = 0$ $d$-$u$ (as in Fig. \ref{fig:N0size}c). First, the three phases (TeS, BS and TrS) in non-chiral shells ($N = 0$) shift to regions of small $u$ in chiral shells ($N = 4$). Second, a new phase, a bent structure (BeS), appears in thicker shells and is greatly favored in inhomogeneous shells. Finally, in thin shells ($N = 4$), as $d$ gets large, TeS is no longer as prevalent as it is in non-chiral shells ($N = 0$), and TrS takes over.

In constrast to non-chiral shells, where the inner drop always move towards the outer boundaries of the drop surface, chiral shells (N = 4) manifest a diversity of energetically favorable positions ($d$) for different shell thickness ($u$). Here, we offer an explanation based on shell chirality $c$. For thin shells, when $c$ is small, similar to non-chiral systems ($c = 0$), the inner drop moves towards the outer drop periphery due to elasticity. As $c$ nears $0.5$, where the director follows a rotation of approximately $\pi$ along the radial direction in uniform shells, the inner drop prefers to stay in the center. Thicker shells ($c \ge 0.75$) adopt configurations similar to those of chiral LC droplets. Because of higher splay and bend distortions on one side of the BeS or RSS droplets, the inner drop favors a slightly off-center position.

\begin{figure*}[h]
\centering
\includegraphics[width=12cm]{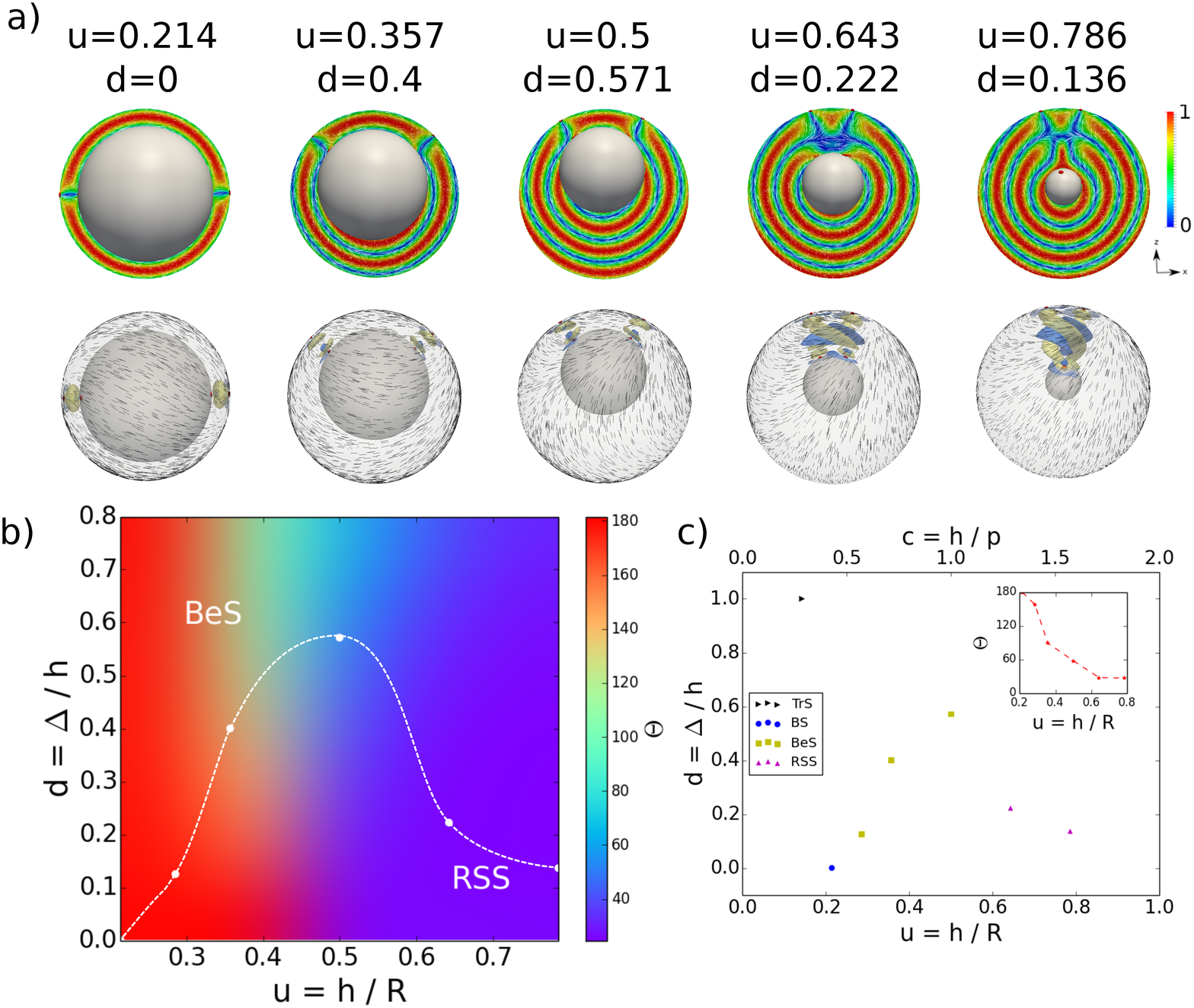}
\caption{(color online) a) Representative shell configurations ($N = 8$) with minimum energies for varying thicknesses. In the first row, the director fields are on $x$-$z$ plane, colored by the projection to $y$-axis. In the second row, director fields are on the outer surface, defects are in red and splay-bend elastic distortions are shown in blue ($S_{\mathrm{SB}} > 0.002$) and in yellow ($S_{\mathrm{SB}} < -0.002$). b) 2D map of $\Theta$ with different $d = \Delta/h$ and $u = h / R$.  c) Diagram of energetically favorable positions $d$ for different shell thickness with $N = 8$. Inset graph shows how $\Theta$ of corresponding configurations varies as $u$ increases.}
\label{fig:N8pos}
\end{figure*}

We close by examining the behavior of highly chiral shells ($N = 8$) in response to $u$ and $d$. As shown in Fig. \ref{fig:N8pos}c, similarly to shells with $N = 4$, when shell chirality $c$ is small, the inner drop moves towards the outer surface; when $c \sim 0.5$, the inner drop prefers to stay in the center; when $c$ further increases, a bent structure appears and the inner drop adopts a slightly off-center position. A unique feature of highly chiral systems is that, when shell chirality is large ($c \ge 0.7$), shells adopt an RSS configuration. The small inner drops in RSS shells prefer to stay near the center position, probably due to a competition between surface anchoring and elastic energy.

When $u \ge 0.214$, chiral shells ($N = 8$) consist of only BS ($\Theta = 180^{\circ}$), BeS ($180^{\circ} > \Theta > 30^{\circ} $) and RSS ($\Theta \sim 30^{\circ}$). One can therefore generate a 2D map of $\Theta$ as a function of $d$ and $u$. For uniform shells ($d = 0$), the BS-to-RSS transition occurs within an extremely narrow range ($\Delta u \le 0.007 $). The sharp response to shell geometry resembles that to chirality, suggesting a potential use of chiral shells as sensors. Deformations of defect configurations occur as shell symmetry is broken. Generally speaking, the shell thickness inhomogeneity induces a decrease of $\Theta$ for BS shells and an increase of $\Theta$ for RSS shells. Interestingly, those minimum energy positions for varying shell thickness form a path with a continuous decrease of $\Theta$ from $180^{\circ}$ to $30^{\circ}$ (as shown in Fig. \ref{fig:N8pos}b), which corresponds to the BS-to-RSS transition. In experiments, the shell thickness $u$ can be easily changed by inducing a difference in osmotic pressure between the inner and outer water phases through addition of salts \cite{lopez2011frustrated}. The transformation path from BS to RSS in highly chiral shells (N = 8) predicted by our simulation (as shown in Fig. \ref{fig:N8pos}a) mimics what is seen in experiments \cite{darmon2015waltzing}. The inset of Fig. \ref{fig:N8pos}c describes how $\Theta$ decreases as shell gets thicker. It demonstrates that when the restriction of a central inner drop is removed, the decrease of $\Theta$ from $180^{\circ}$ to $30^{\circ}$ becomes much slower ($\Delta u \sim 0.43$), compared with the sharp response to shell thickness observed in uniform shells ($\Delta u \le 0.007$).

\section{Conclusion and Prospects}

A systematic study has been presented of non-chiral and chiral nematic LC shells ($R = 1\,\mathrm{\mu m}$) with strong planar degenerate anchoring ($W = 10^{-3}\,\mathrm{J/m^{2}}$). In uniform non-chiral nematic shells, as shell thickness varies, a transition between a bipolar structure (BS) and a tetravalent structure (TeS) occurs near $u^*=0.5$. For asymmetric shells ($\Delta \neq 0$), defect configurations can vary considerably. The energetically favored inner drop positions correspond to locations in the periphery of the outer drop.

For uniform chiral shells ($\Delta = 0$), a phase diagram was presented that includes several novel defect configurations, namely tetravalent structure (TeS), bipolar structure (BS), bent structure (BeS) and radial spherical structure (RSS). To characterize these, a new parameter $c = h / p$ was introduced, which measures chirality in shell geometries. Values of $c = 0.125$ and $c= 0.75$ characterize the TeS-BS and BS-RSS phase boundaries, respectively. It is important to emphasize the sharpness of the BS-RSS transition in response to both chirality and shell geometry, which suggests potential uses for triggerable materials. By examining the elastic forces on the inner drop for intermediate chirality ($N = 4$) and high chirality ($N = 8$), it was shown that the degree of asymmetry of stable configurations for different thicknesses is governed by shell chirality $c$.

Note that in this work the radius $R$ of the drop was fixed, and both surfaces (inner and outer drops) were assumed to exhibit strong degenerate planar anchoring. Many parameters remain to be investigated, including curvature, boundary conditions and temperature. Recent numerical work, for example, has revealed intriguing defect patterns when BPs are confined in slits with thickness comparable to the unit cell size \cite{fukuda2011quasi}. Such systems and others will be considered in a future study.

\begin{acknowledgements}
The authors acknowledge support from the Department of Energy, Basic Energy Sciences, Division of Materials Research, Biomaterials Program under Grant No. DE-SC0004025.
\end{acknowledgements}

\bibliography{ref.bib} 

\begin{thebibliography}{35}%
\makeatletter
\providecommand \@ifxundefined [1]{%
 \@ifx{#1\undefined}
}%
\providecommand \@ifnum [1]{%
 \ifnum #1\expandafter \@firstoftwo
 \else \expandafter \@secondoftwo
 \fi
}%
\providecommand \@ifx [1]{%
 \ifx #1\expandafter \@firstoftwo
 \else \expandafter \@secondoftwo
 \fi
}%
\providecommand \natexlab [1]{#1}%
\providecommand \enquote  [1]{``#1''}%
\providecommand \bibnamefont  [1]{#1}%
\providecommand \bibfnamefont [1]{#1}%
\providecommand \citenamefont [1]{#1}%
\providecommand \href@noop [0]{\@secondoftwo}%
\providecommand \href [0]{\begingroup \@sanitize@url \@href}%
\providecommand \@href[1]{\@@startlink{#1}\@@href}%
\providecommand \@@href[1]{\endgroup#1\@@endlink}%
\providecommand \@sanitize@url [0]{\catcode `\\12\catcode `\$12\catcode
  `\&12\catcode `\#12\catcode `\^12\catcode `\_12\catcode `\%12\relax}%
\providecommand \@@startlink[1]{}%
\providecommand \@@endlink[0]{}%
\providecommand \url  [0]{\begingroup\@sanitize@url \@url }%
\providecommand \@url [1]{\endgroup\@href {#1}{\urlprefix }}%
\providecommand \urlprefix  [0]{URL }%
\providecommand \Eprint [0]{\href }%
\providecommand \doibase [0]{http://dx.doi.org/}%
\providecommand \selectlanguage [0]{\@gobble}%
\providecommand \bibinfo  [0]{\@secondoftwo}%
\providecommand \bibfield  [0]{\@secondoftwo}%
\providecommand \translation [1]{[#1]}%
\providecommand \BibitemOpen [0]{}%
\providecommand \bibitemStop [0]{}%
\providecommand \bibitemNoStop [0]{.\EOS\space}%
\providecommand \EOS [0]{\spacefactor3000\relax}%
\providecommand \BibitemShut  [1]{\csname bibitem#1\endcsname}%
\let\auto@bib@innerbib\@empty
\bibitem [{\citenamefont {Kitzerow}\ and\ \citenamefont
  {Bahr}(2001)}]{kitzerow2001chirality}%
  \BibitemOpen
  \bibfield  {author} {\bibinfo {author} {\bibfnamefont {H.}~\bibnamefont
  {Kitzerow}}\ and\ \bibinfo {author} {\bibfnamefont {C.}~\bibnamefont
  {Bahr}},\ }\href@noop {} {\emph {\bibinfo {title} {Chirality in Liquid
  Crystals}}}\ (\bibinfo  {publisher} {Springer Science \& Business Media},\
  \bibinfo {year} {2001})\BibitemShut {NoStop}%
\bibitem [{\citenamefont
  {Mu{\v{s}}evi{\v{c}}}(2014)}]{muvsevivc2014integrated}%
  \BibitemOpen
  \bibfield  {author} {\bibinfo {author} {\bibfnamefont {I.}~\bibnamefont
  {Mu{\v{s}}evi{\v{c}}}},\ }\href@noop {} {\bibfield  {journal} {\bibinfo
  {journal} {Liq. Cryst.}\ }\textbf {\bibinfo {volume} {41}},\ \bibinfo {pages}
  {418} (\bibinfo {year} {2014})}\BibitemShut {NoStop}%
\bibitem [{\citenamefont {Humar}\ and\ \citenamefont
  {Mu{\v{s}}evi{\v{c}}}(2010)}]{humar20103d}%
  \BibitemOpen
  \bibfield  {author} {\bibinfo {author} {\bibfnamefont {M.}~\bibnamefont
  {Humar}}\ and\ \bibinfo {author} {\bibfnamefont {I.}~\bibnamefont
  {Mu{\v{s}}evi{\v{c}}}},\ }\href@noop {} {\bibfield  {journal} {\bibinfo
  {journal} {Opt. Express}\ }\textbf {\bibinfo {volume} {18}},\ \bibinfo
  {pages} {26995} (\bibinfo {year} {2010})}\BibitemShut {NoStop}%
\bibitem [{\citenamefont {Bezi{\'c}}\ and\ \citenamefont
  {{\v{Z}}umer}(1992)}]{bezic1992structures}%
  \BibitemOpen
  \bibfield  {author} {\bibinfo {author} {\bibfnamefont {J.}~\bibnamefont
  {Bezi{\'c}}}\ and\ \bibinfo {author} {\bibfnamefont {S.}~\bibnamefont
  {{\v{Z}}umer}},\ }\href@noop {} {\bibfield  {journal} {\bibinfo  {journal}
  {Liq. Cryst.}\ }\textbf {\bibinfo {volume} {11}},\ \bibinfo {pages} {593}
  (\bibinfo {year} {1992})}\BibitemShut {NoStop}%
\bibitem [{\citenamefont {Xu}\ and\ \citenamefont
  {Crooker}(1997)}]{xu1997chiral}%
  \BibitemOpen
  \bibfield  {author} {\bibinfo {author} {\bibfnamefont {F.}~\bibnamefont
  {Xu}}\ and\ \bibinfo {author} {\bibfnamefont {P.}~\bibnamefont {Crooker}},\
  }\href@noop {} {\bibfield  {journal} {\bibinfo  {journal} {Phys. Rev. E}\
  }\textbf {\bibinfo {volume} {56}},\ \bibinfo {pages} {6853} (\bibinfo {year}
  {1997})}\BibitemShut {NoStop}%
\bibitem [{\citenamefont {Se{\v{c}}}\ \emph
  {et~al.}(2012{\natexlab{a}})\citenamefont {Se{\v{c}}}, \citenamefont
  {Porenta}, \citenamefont {Ravnik},\ and\ \citenamefont
  {{\v{Z}}umer}}]{sevc2012geometrical}%
  \BibitemOpen
  \bibfield  {author} {\bibinfo {author} {\bibfnamefont {D.}~\bibnamefont
  {Se{\v{c}}}}, \bibinfo {author} {\bibfnamefont {T.}~\bibnamefont {Porenta}},
  \bibinfo {author} {\bibfnamefont {M.}~\bibnamefont {Ravnik}}, \ and\ \bibinfo
  {author} {\bibfnamefont {S.}~\bibnamefont {{\v{Z}}umer}},\ }\href@noop {}
  {\bibfield  {journal} {\bibinfo  {journal} {Soft Matter}\ }\textbf {\bibinfo
  {volume} {8}},\ \bibinfo {pages} {11982} (\bibinfo {year}
  {2012}{\natexlab{a}})}\BibitemShut {NoStop}%
\bibitem [{\citenamefont {Geng}\ \emph {et~al.}(2013)\citenamefont {Geng},
  \citenamefont {Se{\v{c}}}, \citenamefont {Almeida}, \citenamefont
  {Lavrentovich}, \citenamefont {{\v{Z}}umer},\ and\ \citenamefont
  {Godinho}}]{geng2013liquid}%
  \BibitemOpen
  \bibfield  {author} {\bibinfo {author} {\bibfnamefont {Y.}~\bibnamefont
  {Geng}}, \bibinfo {author} {\bibfnamefont {D.}~\bibnamefont {Se{\v{c}}}},
  \bibinfo {author} {\bibfnamefont {P.~L.}\ \bibnamefont {Almeida}}, \bibinfo
  {author} {\bibfnamefont {O.~D.}\ \bibnamefont {Lavrentovich}}, \bibinfo
  {author} {\bibfnamefont {S.}~\bibnamefont {{\v{Z}}umer}}, \ and\ \bibinfo
  {author} {\bibfnamefont {M.~H.}\ \bibnamefont {Godinho}},\ }\href@noop {}
  {\bibfield  {journal} {\bibinfo  {journal} {Soft Matter}\ }\textbf {\bibinfo
  {volume} {9}},\ \bibinfo {pages} {7928} (\bibinfo {year} {2013})}\BibitemShut
  {NoStop}%
\bibitem [{\citenamefont {Se{\v{c}}}\ \emph {et~al.}(2014)\citenamefont
  {Se{\v{c}}}, \citenamefont {{\v{C}}opar},\ and\ \citenamefont
  {{\v{Z}}umer}}]{sevc2014topological}%
  \BibitemOpen
  \bibfield  {author} {\bibinfo {author} {\bibfnamefont {D.}~\bibnamefont
  {Se{\v{c}}}}, \bibinfo {author} {\bibfnamefont {S.}~\bibnamefont
  {{\v{C}}opar}}, \ and\ \bibinfo {author} {\bibfnamefont {S.}~\bibnamefont
  {{\v{Z}}umer}},\ }\href@noop {} {\bibfield  {journal} {\bibinfo  {journal}
  {Nat. Commun.}\ }\textbf {\bibinfo {volume} {5}},\ \bibinfo {pages} {3057}
  (\bibinfo {year} {2014})}\BibitemShut {NoStop}%
\bibitem [{\citenamefont {Orlova}\ \emph {et~al.}(2015)\citenamefont {Orlova},
  \citenamefont {A{\ss}hoff}, \citenamefont {Yamaguchi}, \citenamefont
  {Katsonis},\ and\ \citenamefont {Brasselet}}]{orlova2015creation}%
  \BibitemOpen
  \bibfield  {author} {\bibinfo {author} {\bibfnamefont {T.}~\bibnamefont
  {Orlova}}, \bibinfo {author} {\bibfnamefont {S.~J.}\ \bibnamefont
  {A{\ss}hoff}}, \bibinfo {author} {\bibfnamefont {T.}~\bibnamefont
  {Yamaguchi}}, \bibinfo {author} {\bibfnamefont {N.}~\bibnamefont {Katsonis}},
  \ and\ \bibinfo {author} {\bibfnamefont {E.}~\bibnamefont {Brasselet}},\
  }\href@noop {} {\bibfield  {journal} {\bibinfo  {journal} {Nat. Commun.}\
  }\textbf {\bibinfo {volume} {6}} (\bibinfo {year} {2015})}\BibitemShut
  {NoStop}%
\bibitem [{\citenamefont {Fukuda}\ and\ \citenamefont
  {{\v{Z}}umer}(2011)}]{fukuda2011quasi}%
  \BibitemOpen
  \bibfield  {author} {\bibinfo {author} {\bibfnamefont {J.-i.}\ \bibnamefont
  {Fukuda}}\ and\ \bibinfo {author} {\bibfnamefont {S.}~\bibnamefont
  {{\v{Z}}umer}},\ }\href@noop {} {\bibfield  {journal} {\bibinfo  {journal}
  {Nat. Commun.}\ }\textbf {\bibinfo {volume} {2}},\ \bibinfo {pages} {246}
  (\bibinfo {year} {2011})}\BibitemShut {NoStop}%
\bibitem [{\citenamefont {Wright}\ and\ \citenamefont
  {Mermin}(1989)}]{wright1989crystalline}%
  \BibitemOpen
  \bibfield  {author} {\bibinfo {author} {\bibfnamefont {D.~C.}\ \bibnamefont
  {Wright}}\ and\ \bibinfo {author} {\bibfnamefont {N.~D.}\ \bibnamefont
  {Mermin}},\ }\href@noop {} {\bibfield  {journal} {\bibinfo  {journal} {Rev.
  Mod. phys.}\ }\textbf {\bibinfo {volume} {61}},\ \bibinfo {pages} {385}
  (\bibinfo {year} {1989})}\BibitemShut {NoStop}%
\bibitem [{\citenamefont {Bukusoglu}\ \emph {et~al.}(2015)\citenamefont
  {Bukusoglu}, \citenamefont {Wang}, \citenamefont {Martinez-Gonzalez},
  \citenamefont {de~Pablo},\ and\ \citenamefont
  {Abbott}}]{bukusoglu2015stimuli}%
  \BibitemOpen
  \bibfield  {author} {\bibinfo {author} {\bibfnamefont {E.}~\bibnamefont
  {Bukusoglu}}, \bibinfo {author} {\bibfnamefont {X.}~\bibnamefont {Wang}},
  \bibinfo {author} {\bibfnamefont {J.~A.}\ \bibnamefont {Martinez-Gonzalez}},
  \bibinfo {author} {\bibfnamefont {J.~J.}\ \bibnamefont {de~Pablo}}, \ and\
  \bibinfo {author} {\bibfnamefont {N.~L.}\ \bibnamefont {Abbott}},\
  }\href@noop {} {\bibfield  {journal} {\bibinfo  {journal} {Adv. Mater.
  (Weinheim, Ger.)}\ }\textbf {\bibinfo {volume} {27}},\ \bibinfo {pages}
  {6892} (\bibinfo {year} {2015})}\BibitemShut {NoStop}%
\bibitem [{\citenamefont {Mart{\'\i}nez-Gonz{\'a}lez}\ \emph
  {et~al.}(2015)\citenamefont {Mart{\'\i}nez-Gonz{\'a}lez}, \citenamefont
  {Zhou}, \citenamefont {Rahimi}, \citenamefont {Bukusoglu}, \citenamefont
  {Abbott},\ and\ \citenamefont {de~Pablo}}]{martinez2015blue}%
  \BibitemOpen
  \bibfield  {author} {\bibinfo {author} {\bibfnamefont {J.~A.}\ \bibnamefont
  {Mart{\'\i}nez-Gonz{\'a}lez}}, \bibinfo {author} {\bibfnamefont
  {Y.}~\bibnamefont {Zhou}}, \bibinfo {author} {\bibfnamefont {M.}~\bibnamefont
  {Rahimi}}, \bibinfo {author} {\bibfnamefont {E.}~\bibnamefont {Bukusoglu}},
  \bibinfo {author} {\bibfnamefont {N.~L.}\ \bibnamefont {Abbott}}, \ and\
  \bibinfo {author} {\bibfnamefont {J.~J.}\ \bibnamefont {de~Pablo}},\
  }\href@noop {} {\bibfield  {journal} {\bibinfo  {journal} {Proc. Natl. Acad.
  Sci. U. S. A.}\ ,\ \bibinfo {pages} {201514251}} (\bibinfo {year}
  {2015})}\BibitemShut {NoStop}%
\bibitem [{\citenamefont {Lubensky}\ and\ \citenamefont
  {Prost}(1992)}]{lubensky1992orientational}%
  \BibitemOpen
  \bibfield  {author} {\bibinfo {author} {\bibfnamefont {T.}~\bibnamefont
  {Lubensky}}\ and\ \bibinfo {author} {\bibfnamefont {J.}~\bibnamefont
  {Prost}},\ }\href@noop {} {\bibfield  {journal} {\bibinfo  {journal} {Journal
  de Physique II}\ }\textbf {\bibinfo {volume} {2}},\ \bibinfo {pages} {371}
  (\bibinfo {year} {1992})}\BibitemShut {NoStop}%
\bibitem [{\citenamefont {Nelson}(2002)}]{nelson2002toward}%
  \BibitemOpen
  \bibfield  {author} {\bibinfo {author} {\bibfnamefont {D.~R.}\ \bibnamefont
  {Nelson}},\ }\href@noop {} {\bibfield  {journal} {\bibinfo  {journal} {Nano
  Lett.}\ }\textbf {\bibinfo {volume} {2}},\ \bibinfo {pages} {1125} (\bibinfo
  {year} {2002})}\BibitemShut {NoStop}%
\bibitem [{\citenamefont {Fern{\'a}ndez-Nieves}\ \emph
  {et~al.}(2007)\citenamefont {Fern{\'a}ndez-Nieves}, \citenamefont {Vitelli},
  \citenamefont {Utada}, \citenamefont {Link}, \citenamefont {M{\'a}rquez},
  \citenamefont {Nelson},\ and\ \citenamefont {Weitz}}]{fernandez2007novel}%
  \BibitemOpen
  \bibfield  {author} {\bibinfo {author} {\bibfnamefont {A.}~\bibnamefont
  {Fern{\'a}ndez-Nieves}}, \bibinfo {author} {\bibfnamefont {V.}~\bibnamefont
  {Vitelli}}, \bibinfo {author} {\bibfnamefont {A.~S.}\ \bibnamefont {Utada}},
  \bibinfo {author} {\bibfnamefont {D.~R.}\ \bibnamefont {Link}}, \bibinfo
  {author} {\bibfnamefont {M.}~\bibnamefont {M{\'a}rquez}}, \bibinfo {author}
  {\bibfnamefont {D.~R.}\ \bibnamefont {Nelson}}, \ and\ \bibinfo {author}
  {\bibfnamefont {D.~A.}\ \bibnamefont {Weitz}},\ }\href@noop {} {\bibfield
  {journal} {\bibinfo  {journal} {Phys. Rev. Lett.}\ }\textbf {\bibinfo
  {volume} {99}},\ \bibinfo {pages} {157801} (\bibinfo {year}
  {2007})}\BibitemShut {NoStop}%
\bibitem [{\citenamefont {Lopez-Leon}\ \emph {et~al.}(2011)\citenamefont
  {Lopez-Leon}, \citenamefont {Koning}, \citenamefont {Devaiah}, \citenamefont
  {Vitelli},\ and\ \citenamefont {Fernandez-Nieves}}]{lopez2011frustrated}%
  \BibitemOpen
  \bibfield  {author} {\bibinfo {author} {\bibfnamefont {T.}~\bibnamefont
  {Lopez-Leon}}, \bibinfo {author} {\bibfnamefont {V.}~\bibnamefont {Koning}},
  \bibinfo {author} {\bibfnamefont {K.}~\bibnamefont {Devaiah}}, \bibinfo
  {author} {\bibfnamefont {V.}~\bibnamefont {Vitelli}}, \ and\ \bibinfo
  {author} {\bibfnamefont {A.}~\bibnamefont {Fernandez-Nieves}},\ }\href@noop
  {} {\bibfield  {journal} {\bibinfo  {journal} {Nat. Phys.}\ }\textbf
  {\bibinfo {volume} {7}},\ \bibinfo {pages} {391} (\bibinfo {year}
  {2011})}\BibitemShut {NoStop}%
\bibitem [{\citenamefont {Koning}\ \emph {et~al.}(2015)\citenamefont {Koning},
  \citenamefont {Lopez-Leon}, \citenamefont {Darmon}, \citenamefont
  {Fernandez-Nieves},\ and\ \citenamefont {Vitelli}}]{koning2015spherical}%
  \BibitemOpen
  \bibfield  {author} {\bibinfo {author} {\bibfnamefont {V.}~\bibnamefont
  {Koning}}, \bibinfo {author} {\bibfnamefont {T.}~\bibnamefont {Lopez-Leon}},
  \bibinfo {author} {\bibfnamefont {A.}~\bibnamefont {Darmon}}, \bibinfo
  {author} {\bibfnamefont {A.}~\bibnamefont {Fernandez-Nieves}}, \ and\
  \bibinfo {author} {\bibfnamefont {V.}~\bibnamefont {Vitelli}},\ }\href@noop
  {} {\bibfield  {journal} {\bibinfo  {journal} {arXiv preprint
  arXiv:1502.03742}\ } (\bibinfo {year} {2015})}\BibitemShut {NoStop}%
\bibitem [{\citenamefont {Lopez-Leon}\ and\ \citenamefont
  {Fernandez-Nieves}(2011)}]{lopez2011drops}%
  \BibitemOpen
  \bibfield  {author} {\bibinfo {author} {\bibfnamefont {T.}~\bibnamefont
  {Lopez-Leon}}\ and\ \bibinfo {author} {\bibfnamefont {A.}~\bibnamefont
  {Fernandez-Nieves}},\ }\href@noop {} {\bibfield  {journal} {\bibinfo
  {journal} {Colloid and Polym. Sci.}\ }\textbf {\bibinfo {volume} {289}},\
  \bibinfo {pages} {345} (\bibinfo {year} {2011})}\BibitemShut {NoStop}%
\bibitem [{\citenamefont {Lopez-Leon}\ \emph {et~al.}(2012)\citenamefont
  {Lopez-Leon}, \citenamefont {Bates},\ and\ \citenamefont
  {Fernandez-Nieves}}]{lopez2012defect}%
  \BibitemOpen
  \bibfield  {author} {\bibinfo {author} {\bibfnamefont {T.}~\bibnamefont
  {Lopez-Leon}}, \bibinfo {author} {\bibfnamefont {M.~A.}\ \bibnamefont
  {Bates}}, \ and\ \bibinfo {author} {\bibfnamefont {A.}~\bibnamefont
  {Fernandez-Nieves}},\ }\href@noop {} {\bibfield  {journal} {\bibinfo
  {journal} {Phys. Rev. E}\ }\textbf {\bibinfo {volume} {86}},\ \bibinfo
  {pages} {030702} (\bibinfo {year} {2012})}\BibitemShut {NoStop}%
\bibitem [{\citenamefont {Lopez-Leon}\ and\ \citenamefont
  {Fernandez-Nieves}(2009)}]{lopez2009topological}%
  \BibitemOpen
  \bibfield  {author} {\bibinfo {author} {\bibfnamefont {T.}~\bibnamefont
  {Lopez-Leon}}\ and\ \bibinfo {author} {\bibfnamefont {A.}~\bibnamefont
  {Fernandez-Nieves}},\ }\href@noop {} {\bibfield  {journal} {\bibinfo
  {journal} {Phys. Rev. E}\ }\textbf {\bibinfo {volume} {79}},\ \bibinfo
  {pages} {021707} (\bibinfo {year} {2009})}\BibitemShut {NoStop}%
\bibitem [{\citenamefont {Vitelli}\ and\ \citenamefont
  {Nelson}(2006)}]{vitelli2006nematic}%
  \BibitemOpen
  \bibfield  {author} {\bibinfo {author} {\bibfnamefont {V.}~\bibnamefont
  {Vitelli}}\ and\ \bibinfo {author} {\bibfnamefont {D.~R.}\ \bibnamefont
  {Nelson}},\ }\href@noop {} {\bibfield  {journal} {\bibinfo  {journal} {Phys.
  Rev. E}\ }\textbf {\bibinfo {volume} {74}},\ \bibinfo {pages} {021711}
  (\bibinfo {year} {2006})}\BibitemShut {NoStop}%
\bibitem [{\citenamefont {Shin}\ \emph {et~al.}(2008)\citenamefont {Shin},
  \citenamefont {Bowick},\ and\ \citenamefont {Xing}}]{shin2008topological}%
  \BibitemOpen
  \bibfield  {author} {\bibinfo {author} {\bibfnamefont {H.}~\bibnamefont
  {Shin}}, \bibinfo {author} {\bibfnamefont {M.~J.}\ \bibnamefont {Bowick}}, \
  and\ \bibinfo {author} {\bibfnamefont {X.}~\bibnamefont {Xing}},\ }\href@noop
  {} {\bibfield  {journal} {\bibinfo  {journal} {Phys. Rev. Lett.}\ }\textbf
  {\bibinfo {volume} {101}},\ \bibinfo {pages} {037802} (\bibinfo {year}
  {2008})}\BibitemShut {NoStop}%
\bibitem [{\citenamefont {Se{\v{c}}}\ \emph
  {et~al.}(2012{\natexlab{b}})\citenamefont {Se{\v{c}}}, \citenamefont
  {Lopez-Leon}, \citenamefont {Nobili}, \citenamefont {Blanc}, \citenamefont
  {Fernandez-Nieves}, \citenamefont {Ravnik},\ and\ \citenamefont
  {{\v{Z}}umer}}]{sevc2012defect}%
  \BibitemOpen
  \bibfield  {author} {\bibinfo {author} {\bibfnamefont {D.}~\bibnamefont
  {Se{\v{c}}}}, \bibinfo {author} {\bibfnamefont {T.}~\bibnamefont
  {Lopez-Leon}}, \bibinfo {author} {\bibfnamefont {M.}~\bibnamefont {Nobili}},
  \bibinfo {author} {\bibfnamefont {C.}~\bibnamefont {Blanc}}, \bibinfo
  {author} {\bibfnamefont {A.}~\bibnamefont {Fernandez-Nieves}}, \bibinfo
  {author} {\bibfnamefont {M.}~\bibnamefont {Ravnik}}, \ and\ \bibinfo {author}
  {\bibfnamefont {S.}~\bibnamefont {{\v{Z}}umer}},\ }\href@noop {} {\bibfield
  {journal} {\bibinfo  {journal} {Phys. Rev. E}\ }\textbf {\bibinfo {volume}
  {86}},\ \bibinfo {pages} {020705} (\bibinfo {year}
  {2012}{\natexlab{b}})}\BibitemShut {NoStop}%
\bibitem [{\citenamefont {Seyednejad}\ \emph {et~al.}(2013)\citenamefont
  {Seyednejad}, \citenamefont {Mozaffari},\ and\ \citenamefont
  {Ejtehadi}}]{seyednejad2013confined}%
  \BibitemOpen
  \bibfield  {author} {\bibinfo {author} {\bibfnamefont {S.~R.}\ \bibnamefont
  {Seyednejad}}, \bibinfo {author} {\bibfnamefont {M.~R.}\ \bibnamefont
  {Mozaffari}}, \ and\ \bibinfo {author} {\bibfnamefont {M.~R.}\ \bibnamefont
  {Ejtehadi}},\ }\href@noop {} {\bibfield  {journal} {\bibinfo  {journal}
  {Phys. Rev. E}\ }\textbf {\bibinfo {volume} {88}},\ \bibinfo {pages} {012508}
  (\bibinfo {year} {2013})}\BibitemShut {NoStop}%
\bibitem [{\citenamefont {Wand}\ and\ \citenamefont
  {Bates}(2015)}]{wand2015monte}%
  \BibitemOpen
  \bibfield  {author} {\bibinfo {author} {\bibfnamefont {C.~R.}\ \bibnamefont
  {Wand}}\ and\ \bibinfo {author} {\bibfnamefont {M.~A.}\ \bibnamefont
  {Bates}},\ }\href@noop {} {\bibfield  {journal} {\bibinfo  {journal} {Phys.
  Rev. E}\ }\textbf {\bibinfo {volume} {91}},\ \bibinfo {pages} {012502}
  (\bibinfo {year} {2015})}\BibitemShut {NoStop}%
\bibitem [{\citenamefont {Gharbi}\ \emph {et~al.}(2013)\citenamefont {Gharbi},
  \citenamefont {Se{\v{c}}}, \citenamefont {Lopez-Leon}, \citenamefont
  {Nobili}, \citenamefont {Ravnik}, \citenamefont {{\v{Z}}umer},\ and\
  \citenamefont {Blanc}}]{gharbi2013microparticles}%
  \BibitemOpen
  \bibfield  {author} {\bibinfo {author} {\bibfnamefont {M.~A.}\ \bibnamefont
  {Gharbi}}, \bibinfo {author} {\bibfnamefont {D.}~\bibnamefont {Se{\v{c}}}},
  \bibinfo {author} {\bibfnamefont {T.}~\bibnamefont {Lopez-Leon}}, \bibinfo
  {author} {\bibfnamefont {M.}~\bibnamefont {Nobili}}, \bibinfo {author}
  {\bibfnamefont {M.}~\bibnamefont {Ravnik}}, \bibinfo {author} {\bibfnamefont
  {S.}~\bibnamefont {{\v{Z}}umer}}, \ and\ \bibinfo {author} {\bibfnamefont
  {C.}~\bibnamefont {Blanc}},\ }\href@noop {} {\bibfield  {journal} {\bibinfo
  {journal} {Soft Matter}\ }\textbf {\bibinfo {volume} {9}},\ \bibinfo {pages}
  {6911} (\bibinfo {year} {2013})}\BibitemShut {NoStop}%
\bibitem [{\citenamefont {De~Gennes}\ and\ \citenamefont
  {Prost}(1993)}]{de1993physics}%
  \BibitemOpen
  \bibfield  {author} {\bibinfo {author} {\bibfnamefont {P.-G.}\ \bibnamefont
  {De~Gennes}}\ and\ \bibinfo {author} {\bibfnamefont {J.}~\bibnamefont
  {Prost}},\ }\href@noop {} {\emph {\bibinfo {title} {The Physics of Liquid
  Crystals}}},\ Vol.~\bibinfo {volume} {23}\ (\bibinfo  {publisher} {Clarendon
  press Oxford},\ \bibinfo {year} {1993})\BibitemShut {NoStop}%
\bibitem [{\citenamefont {Kl{\'e}man}\ and\ \citenamefont
  {Laverntovich}(2007)}]{kleman2007soft}%
  \BibitemOpen
  \bibfield  {author} {\bibinfo {author} {\bibfnamefont {M.}~\bibnamefont
  {Kl{\'e}man}}\ and\ \bibinfo {author} {\bibfnamefont {O.~D.}\ \bibnamefont
  {Laverntovich}},\ }\href@noop {} {\emph {\bibinfo {title} {Soft Matter
  Physics: an Introduction}}}\ (\bibinfo  {publisher} {Springer Science \&
  Business Media},\ \bibinfo {year} {2007})\BibitemShut {NoStop}%
\bibitem [{\citenamefont {Ravnik}\ and\ \citenamefont
  {{\v{Z}}umer}(2009)}]{ravnik2009landau}%
  \BibitemOpen
  \bibfield  {author} {\bibinfo {author} {\bibfnamefont {M.}~\bibnamefont
  {Ravnik}}\ and\ \bibinfo {author} {\bibfnamefont {S.}~\bibnamefont
  {{\v{Z}}umer}},\ }\href@noop {} {\bibfield  {journal} {\bibinfo  {journal}
  {Liq. Cryst.}\ }\textbf {\bibinfo {volume} {36}},\ \bibinfo {pages} {1201}
  (\bibinfo {year} {2009})}\BibitemShut {NoStop}%
\bibitem [{\citenamefont {{\v{C}}opar}\ \emph {et~al.}(2013)\citenamefont
  {{\v{C}}opar}, \citenamefont {Porenta},\ and\ \citenamefont
  {{\v{Z}}umer}}]{vcopar2013visualisation}%
  \BibitemOpen
  \bibfield  {author} {\bibinfo {author} {\bibfnamefont {S.}~\bibnamefont
  {{\v{C}}opar}}, \bibinfo {author} {\bibfnamefont {T.}~\bibnamefont
  {Porenta}}, \ and\ \bibinfo {author} {\bibfnamefont {S.}~\bibnamefont
  {{\v{Z}}umer}},\ }\href@noop {} {\bibfield  {journal} {\bibinfo  {journal}
  {Liq. Cryst.}\ }\textbf {\bibinfo {volume} {40}},\ \bibinfo {pages} {1759}
  (\bibinfo {year} {2013})}\BibitemShut {NoStop}%
\bibitem [{\citenamefont {Utada}\ \emph {et~al.}(2005)\citenamefont {Utada},
  \citenamefont {Lorenceau}, \citenamefont {Link}, \citenamefont {Kaplan},
  \citenamefont {Stone},\ and\ \citenamefont {Weitz}}]{utada2005monodisperse}%
  \BibitemOpen
  \bibfield  {author} {\bibinfo {author} {\bibfnamefont {A.}~\bibnamefont
  {Utada}}, \bibinfo {author} {\bibfnamefont {E.}~\bibnamefont {Lorenceau}},
  \bibinfo {author} {\bibfnamefont {D.}~\bibnamefont {Link}}, \bibinfo {author}
  {\bibfnamefont {P.}~\bibnamefont {Kaplan}}, \bibinfo {author} {\bibfnamefont
  {H.}~\bibnamefont {Stone}}, \ and\ \bibinfo {author} {\bibfnamefont
  {D.}~\bibnamefont {Weitz}},\ }\href@noop {} {\bibfield  {journal} {\bibinfo
  {journal} {Science}\ }\textbf {\bibinfo {volume} {308}},\ \bibinfo {pages}
  {537} (\bibinfo {year} {2005})}\BibitemShut {NoStop}%
\bibitem [{\citenamefont {Poincar{\'e}}(1885)}]{poincare1885j}%
  \BibitemOpen
  \bibfield  {author} {\bibinfo {author} {\bibfnamefont {H.}~\bibnamefont
  {Poincar{\'e}}},\ }\href@noop {} {\bibfield  {journal} {\bibinfo  {journal}
  {Journal de Mathématiques Pures et Appliquées}\ }\textbf {\bibinfo {volume}
  {1}},\ \bibinfo {pages} {167} (\bibinfo {year} {1885})}\BibitemShut {NoStop}%
\bibitem [{\citenamefont {Zhou}\ \emph {et~al.}(2016)\citenamefont {Zhou},
  \citenamefont {Bukusoglu}, \citenamefont {Mart{\'\i}nez-Gonz{\'a}lez},
  \citenamefont {Rahimi}, \citenamefont {Roberts}, \citenamefont {Zhang},
  \citenamefont {Wang}, \citenamefont {Abbott},\ and\ \citenamefont
  {de~Pablo}}]{zhou2016structural}%
  \BibitemOpen
  \bibfield  {author} {\bibinfo {author} {\bibfnamefont {Y.}~\bibnamefont
  {Zhou}}, \bibinfo {author} {\bibfnamefont {E.}~\bibnamefont {Bukusoglu}},
  \bibinfo {author} {\bibfnamefont {J.~A.}\ \bibnamefont
  {Mart{\'\i}nez-Gonz{\'a}lez}}, \bibinfo {author} {\bibfnamefont
  {M.}~\bibnamefont {Rahimi}}, \bibinfo {author} {\bibfnamefont {T.~F.}\
  \bibnamefont {Roberts}}, \bibinfo {author} {\bibfnamefont {R.}~\bibnamefont
  {Zhang}}, \bibinfo {author} {\bibfnamefont {X.}~\bibnamefont {Wang}},
  \bibinfo {author} {\bibfnamefont {N.~L.}\ \bibnamefont {Abbott}}, \ and\
  \bibinfo {author} {\bibfnamefont {J.~J.}\ \bibnamefont {de~Pablo}},\
  }\href@noop {} {\bibfield  {journal} {\bibinfo  {journal} {ACS nano}\ }
  (\bibinfo {year} {2016})}\BibitemShut {NoStop}%
\bibitem [{\citenamefont {Darmon}\ \emph {et~al.}(2015)\citenamefont {Darmon},
  \citenamefont {Benzaquen}, \citenamefont {Dauchot},\ and\ \citenamefont
  {Lopez-Leon}}]{darmon2015waltzing}%
  \BibitemOpen
  \bibfield  {author} {\bibinfo {author} {\bibfnamefont {A.}~\bibnamefont
  {Darmon}}, \bibinfo {author} {\bibfnamefont {M.}~\bibnamefont {Benzaquen}},
  \bibinfo {author} {\bibfnamefont {O.}~\bibnamefont {Dauchot}}, \ and\
  \bibinfo {author} {\bibfnamefont {T.}~\bibnamefont {Lopez-Leon}},\
  }\href@noop {} {\bibfield  {journal} {\bibinfo  {journal} {arXiv preprint
  arXiv:1512.06039}\ } (\bibinfo {year} {2015})}\BibitemShut {NoStop}%
\end{thebibliography}%

\end{document}